\documentclass[11pt]{article}
\usepackage{latexsym}
\usepackage{graphicx}
\usepackage{amsfonts,amssymb}

\begin{document}

\title{Evolving Model of Weighted Networks Inspired by Scientific Collaboration Networks}% Force line breaks with \\

\author{Menghui Li$^{1}$, Jinshan Wu$^{2}$, Dahui Wang$^{1}$, Tao Zhou$^{3}$, Zengru Di$^{1}$,
Ying Fan$^{1}$\footnote{Author for correspondence:
yfan@bnu.edu.cn}\\\\ 1. Department of Systems Science, School of Management,\\
 Beijing Normal University, Beijing 100875, P.R.China.\\
2. Department of Physics \& Astronomy, University of British
Columbia,\\ Vancouver, B.C. Canada, V6T 1Z1.\\
3. Department of Modern Physics, University of Science and
Technology of China,\\ Hefei Anhui, 230026, P.R.China}

\maketitle

\begin{abstract}
Inspired by scientific collaboration networks, especially our
empirical analysis of the network of econophysicists, an
evolutionary model for weighted networks is proposed. Both
degree-driven and weight-driven models are considered. Compared
with the BA model and other evolving models with preferential
attachment, there are two significant generalizations. First,
besides the new vertex added in at every time step, old vertices
can also attempt to build up new links, or to reconnect the
existing links. The reconnection between both new-old and old-old
nodes are recorded and the connecting times on every link is
converted into the weight of the link. This provides a natural way
for the evolution of edge weight. Second, besides degree and the
weight of vertices, a path-related local information is also used
as a reference in the preferential attachment. The path-related
preferential attachment mechanism significantly increases the
clustering coefficient of the network. The model shows the
scale-free phenomena in degree and weight distribution. It also
gives well qualitatively consistent behavior with the empirical
results.
\end{abstract}

Pacs: 89.75.Hc 05.40.-a 87.23.Kg

\section{\label{Introduction}Introduction}

Network analysis is now widely used in many fields \cite{Review1,
Review2}. Recently more and more works on weighted networks
appears in both empirical and modelling analysis. In a weighted
network, the weight on the edges provides a natural way to take
into account the interaction strength, while in a binary network,
the edges only represents the presence or absence of interaction.
This capability will probably carry more information about the
interaction.

The first problem before any analysis can be applied to the
weighted networks is how to assign the weight to edges. This
problem is quite non-trivial. Several ways to assign the weight
have been introduced. One is to transfer some quantities from
non-weighted networks into the weight of edges. They are usually
related to the degree or other intrinsic quantities of the
nodes\cite{Yook,Zheng,Macdonald}. As in \cite{Macdonald}, the
weight of an edge is measured by the point degree of its two ends,
which are defined phenomenologically from binary networks. It is
helpful to describe new properties of the binary networks. but it
includes no more information than the origin binary networks.
Sometimes, the real-world phenomena investigated provide a
typically natural measurement of the weight, such as the number of
flights or seats between any two cities in airport
networks\cite{W.Li,Barrat1,Bagler}, the reaction rate in metabolic
networks\cite{Nature} and so on. In the works of modelling
weighted networks, weights on edges are generated from priori
distribution\cite{Antal,Goh2,Park}. From the view point of
empirical study, we never know such models already acquire the
real structure of weighted networks or not.

However, some weighted networks such as scientific collaboration
networks are different with the above networks. In the
collaboration networks, the connection times is a natural quantity
which is related very closely to the weight. But there is no
explicitly expression between this quantity and weight. Let's
think about the times of coauthoring between two scientists.
Obviously, more times represents closer relationship in the sense
of transportation of scientific ideas. Therefore, in scientific
collaboration networks, usually the happening times of the event
is converted as the weight of the edge. Yet different authors may
use different expressions\cite{Barrat1,Newman1,Newman2,Fan,Li}. As
to which definition behaviors better, and whether or not there are
some general rules to define weight, we do not have the final
answer yet.

The second problem related to the weighted networks is how to
extract information from weighted networks constructed by the
above ways. Especially one may concern about what's the role of
weight, or what's the significant difference brought by weights
compared with binary networks. In order to answer the above
questions raised from those two aspects, we have to consider the
third problem, modelling the weighted networks.

For instance, by investigation of modelling works, if we find that
in order to construct a well-behavior model of weighted networks,
the degree is the only variable directly coupled with evolution,
while the weight is never needed to directly be brought into the
evolutionary process, then we may think that the weight just
relies on a higher level structure. The weight is not crucial in
network analysis although it is important and necessary. Or quite
the contrary, if in order to get weighted network behaviors
consistently with real phenomena in the modelling work, the weight
must be coupled directly with the evolution. Then the weight
should play a significant role in the way to extract information
from weighted networks.

In this paper, we tried both degree-referred preferential
attachment and weight-referred preferential attachment in our
evolving model and compared the results with the empirical
analysis from \cite{Newman2,Fan,Li}.

Actually, there are already many evolving models for weighted
networks. Some models introduced prior weights into edges with the
evolution of networks. In \cite{Yook}, each link $j\leftrightarrow
i$ from the newly added node $j$ is assigned a weight as
$w_{ji}=\frac{k_i}{\sum_{\{i^{\prime}\}}k_{i^{\prime}}}$, where
$\{i^{\prime}\}$ represents a sum over the $m$ existing nodes to
which the new node $j$ is connected. Zheng\cite{Zheng} has
improved this idea. In his model, the weight of a link depends not
only on total degree of the existing nodes, but also on some
intrinsic quality ("fittness") of the nodes. In \cite{W.Je}, the
weight of a link depends on randomly modified intervals between
the time at which linked vertices are connected to the system. In
\cite{Park}, the weight $w_{ij}$ of a link $l_{ij}$ connecting a
pair of nodes ($i$ and $j$) is defined as $w_{ij} =(w_i+w_j)/2$,
and $w_i$ is defined as $i$ node's assigned number (from 1 to $N$)
divided by $N$. In some evolving models\cite{Antal,Goh2}, the
weight $w$ is assigned to the link when it is created and it is
drawn from a certain distribution. As pointed out in
\cite{Barrat}, most models here are not really evolutionary models
in the sense of weight. The weight keeps the same value after it
was assigned onto its edge. Or some extra quantities are
introduced to drive the evolution of networks.

Recently, some evolving models are set up in which the weights are
coupled directly with the network evolution. In the
paper\cite{Barrat}, a weight-driven model was proposed and the
weight of link changes with the network evolution. In this paper,
the new edges starting from the new vertex added in at every time
step are preferentially attached to old vertices determined by
their strength, or vertex weight. After the attachment, an
increase of weight $\delta$ is distributed among all the edges
connecting to the chosen old vertices. The model yields a
nontrivial time evolution of vertices' properties and scale-free
behavior for the weight, strength, and degree distributions. In
the paper\cite{Bianconi}, Bianconi has presented a model with
co-evolution of link weight and strength. In his weighted fitness
network model, the fitness of node and link and results in the
structural phase transition of the network are introduced. In
\cite{W.Je2}, the network evolves with connectivity-driven
topology and with the weight assigned from a special distribution
$\rho_k(x)$ of weights.

Although the models mentioned above coupled the weight and network
evolution, we think that the dynamical process of the weight in
Barrat's model\cite{Barrat} is quite artificial or say not very
general, or like in the other two, extra quantities not rooted in
network has to be used. The authors of \cite{Barrat} gave some
arguments for this as to justify the process from the background
of airport network\cite{Barrat}. But they took weight as a
quantity independent on connecting. However, as we have mentioned
before, weight usually related closely to connecting times.
Especially for the actors and scientists collaboration networks,
using weight converted from connecting times is a convenient way
to construct weighted networks. Therefore, it seems that such a
pure weight-driven model depends too much on this artificial
dynamical process of weight. Now, our empirical investigation on
scientific collaboration networks give us some hints on modelling
weighted networks.

In our model, we keep the relationship between weight and
connecting times, and only quantities directly rooted in networks
are used. So the picture of the evolution looks like the
connecting times evolve according to weight, and then the new
connecting times comes into the weight, which drives the evolution
of the system again. Or in our degree-driven model, connecting
times evolve according to degree, and degree increase due to
connecting, and then all the connecting times are recorded and
converted into weight.

Another important improvement of our model is the introduction of
local-path-related preferential attachment, the $\delta$ term in
our model. This mechanism works for the network evolution in the
real world but is neglected by other models. It is helpful to
increase the clustering coefficient of the networks. One major
difference between empirical results and most models is about the
clustering coefficient. Usually, BA model\cite{BAmodel} or similar
models\cite{Barrat}, given a quite low clustering coefficient
while in reality, real phenomena show highly clustered behavior.
Of course, the WS model of small world network\cite{WS} gives high
clustering coefficient because it starts from a regular network,
not on the way of evolutionary network models. Some evolutionary
models do give high clustering
coefficient\cite{Holme,Davidsen,Barabasi}. In \cite{Holme}, if an
edge between $v$ and $w$ was added, then add one more edge from
$v$ to a randomly chosen neighbor of $w$. In \cite{Davidsen}, one
randomly chosen person introduces two random acquaintances to each
another who haven't met before. Another idea is to introduce an
extra Euclidean distance, and vertices prefer to interact with
nearby vertices. Therefore, in order to increase the clustering
coefficient, new mechanisms which are not rooted directly in the
network have been introduced. Now, we introduce the $\delta$
mechanism, where all quantities still comes directly from the
topological structure. This requires no more extra information,
but just a little knowledge about the local structure. Here `a
little' means one only need to know the information about the
second, or third nearest neighbors, not any more.

The detailed comparison will be done between the results from the
models and our empirical results from \cite{Fan, Li}. The
description of the general model is given in Section
$\S$\ref{model}. The asymptotic distributions of vertex weights
for the weight-driven case is also given analytically in Section
$\S$\ref{Simulation}, and they are well consistent with results of
numerical simulations. In Section $\S$\ref{extend}, in order to
compare with the empirical study of econophysists collaboration
network, we extend our model onto directed weighted networks. In
this comparison, they show nice agreement.

\section{\label{model}Models and theoretical analysis}

\subsection{The model}
A $N$-vertex weighted network is defined by a $N\times N$ matrix
$w_{ij}$, which represents the weight on the edge from vertex $i$
to $j$. Similarity weight is used here. So the larger the weight
is, the closer the relation between the two ends nodes are.
$w_{lm}=0$ means no relation between vertex $l$ and $m$. Suppose
the edge weight $w_{ij}$ is related to the connecting times
$T_{ij}$ between vertex $i$ and $j$, by
\begin{equation}
w_{ij} = f\left(T_{ij}\right),
\end{equation}
such as the tanh function $w_{ij} = \tanh\left(\alpha
T_{ij}\right)$ we used in \cite{Fan,Li}, or just linear relation
$w_{ij} = \alpha T_{ij}$ used by other
authors\cite{Barrat1,Newman2}.

Our most general model is given as follow. Starting from a fully
connected $n_{0}$ initial network, with initial times $T_{ij} = 1$
(and initial weight $w_{ij} = f\left(1\right)$), at every time
step,
\begin{enumerate}
\item One new vertex is added into this network, and $l$ old
vertices are randomly chosen from the existing network.

\item Every one (denoted as vertex $n$) of them can initially
activate a temptation to build up $m$ connections. The probability
for every link from $n$ connecting onto vertex $i$ is given by
\begin{equation}
\Pi_{n\rightarrow i} = \left(1-p\right)\frac{k_{i}}{\sum_{j}k_{j}}
+ \left(p-\delta\right)\frac{w_{i}}{\sum_{j}w_{j}} + \delta
\frac{l_{ni}}{\sum_{j\in\partial^{1,2}_{n}}l_{nj}}, \label{prob}
\end{equation}
where $k_{i}$ is the degree of vertex $i$, $w_{i} =
\sum_{j}w_{ji}$ is the ``onto'' vertex weight of vertex $i$,
$l_{ni}$ is the similarity distance\cite{Li} from $n$ to $i$, and
$\partial^{d}_{n}$ means the $d$th neighbors of vertex $n$. For
example, $\partial^{1}_{n}$ is the set of nearest neighbors,
$\partial^{2}_{n}$ means the second nearest neighbors, so that
$\partial^{1,2}_{n}$ in the expression refers to both of them.
Intuitively, similarity distance means the maximum distance
between two vertices because the weight is defined here in the way
that the larger the closer. Usually, in calculation of network
analysis, the dissimilarity distance corresponding to the shortest
distance is used more often.

\item After we got an end node $i^{*}$ chosen from all vertices
over the existing network by the probability above in
equ(\ref{prob}), the connecting times between vertex $n$ and
$i^{*}$ increased by
\begin{equation}
T_{ni^{*}}\left(t+1\right) = T_{ni^{*}}\left(t\right) + 1.
\end{equation}
\item The weight of the edges changes as
\begin{equation}
w_{ni^{*}}\left(t+1\right) =
f\left(T_{ni^{*}}\left(t+1\right)\right).
\end{equation}
\end{enumerate}
Although our general model defined above can be applied to
directed networks, in the following analysis we assume that
$w_{ij} = w_{ji}$. An increase on $T_{ij}$ immediately reflects
another increase on $T_{ji}$. Except for the comparison with
empirical results, on most cases, the linear function is used for
the relationship between connecting times and weight for the
simplicity,
\begin{equation}
w_{ij} = \alpha T_{ij}. \label{linkweight}
\end{equation}

\subsection{Analytic results of the weight distribution}
\label{analytic}

Now we try to get the analytical results for the vertex weight
distribution under the simplest weight-driven model. For the link
weight given by equ(\ref{linkweight}), the weight of vertex is
given by
\begin{equation}
w_{i} = \sum_{j} T_{ji} \label{strength}
\end{equation}
we suppose that the newly added vertex and the old vertices are
informed of the weight of the other vertices and the network is
pure weight-driven. In this case it is attached with preferential
linking described by $p=1$ and $\delta=0$ in equ(\ref{prob}), that
is the the connection probability is
\begin{equation}
\Pi_{n\rightarrow i} =\frac{w_{i}}{\sum_{j}w_{j}}
\label{weight-driven}
\end{equation}
The master equation for the evolution of the average number of
vertices with weight $w$ at time $t$ is
\begin{eqnarray}
\begin{array}{llll}
N(w,t+1) & = N(w,t) & + & m\cdot(1+l)\cdot \frac{(w-1)\cdot
N(w-1,t)-w\cdot N(w,t)}{\sum_{w}w\cdot N(w,t)}\\
& &- &\frac{l\cdot N(w,t)}{N} + \frac{l\cdot N(w-m,t)}{N}
+\delta_{w,m}
\end{array}\label{master}
\end{eqnarray}

\begin{figure}
\center \includegraphics[width=7cm]{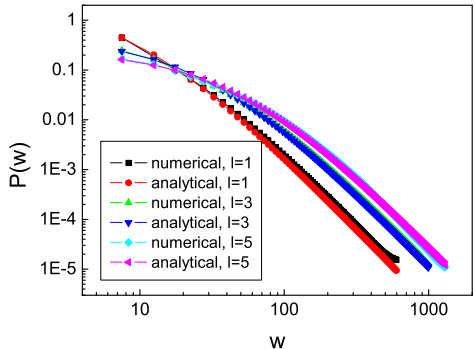}
\caption{Comparison between the numerical solution results from
eq(\ref{master}) and the analytical results of eq(\ref{final}) for
different $l$. Other parameters are $n_{0}=10$, $m=5$.}
\label{EquationLine}
\end{figure}

Here $\sum_{w}w\cdot N(w,t-1)=2\cdot E_0+2\cdot m\cdot (1+l)\cdot
t$ is the total weight and $N=n_0+t$ is the size of system at time
$t$. The equation describes the increasing of preferential linking
since the new vertex is added and old vertices are selected. The
first term reflects the preferential attachment
(\ref{weight-driven}) used to select the other end of the link,
while the following two items correspond to the random selection
of $l$ old vertices. When $E_0$ and $n_0$ are much smaller than
$t$, the size of the network $N$ is approximately the time steps
$t$. Then the master equation(\ref{master}) can be written as
\begin{equation}
\begin{array}{l}
(t+1)\cdot p(w,t+1)=t\cdot p(w,t)+ \frac{1}{2}\cdot[(w-1)\cdot
p(w-1,t)\\
\\
-k\cdot p(w,t)]-l\cdot p(w,t)+l\cdot p(w-m,t)+\delta_{w,m}
\end{array}\label{ppp}
\end{equation}
where $p(w,t)\simeq \frac{N(w,t)}{t}$ is the density of vertices
with strength $w$ at time $t$\cite{Oxford}. When $t$ is
larger($t\gg 1$) enough,
\begin{equation}
(t+1)\cdot p(w,t+1)-t\cdot p(w,t)=p(w,t).
\end{equation}
We get from equ(\ref{ppp})
\begin{equation}
\begin{array}{l}
p(w)=\frac{-d(wp(w))}{2\cdot dw}-l\cdot [p(w)-p(w-m)]+\delta_{w,m}\\
\\
=\frac{-d(wp(w))}{2\cdot dw}-l\cdot [p(w)-p(w-1)+p(w-1)
-p(w-2)\\
\\
+p(w-2)-\cdots+p(w-m+1)-p(w-m)]+\delta_{w,m}\\
\\
=\frac{-d(wp(w))}{2\cdot dw}-l\cdot m\cdot
\frac{dp(w)}{dw}+\delta_{w,m}
\end{array}
\end{equation}
This is further written as
\begin{equation}
p(w)+\frac{d}{dw}[\frac{1}{2}\cdot w\cdot p(w)+l\cdot m\cdot
p(w)]=\delta_{w,m}
\end{equation}
For $w\neq m$, we get
\begin{equation}
(2\cdot l\cdot m+w)\cdot \frac{dp(w)}{dw}=-3\cdot p(w)
\end{equation}
We arrive at the final vertex weight distribution
\begin{equation}
p(w)\propto (w+2\cdot l\cdot m)^{-3} \label{final}
\end{equation}
In Fig(\ref{EquationLine}), we compare the numerical solution of
equ(\ref{master}) with the analytical results equ(\ref{final}), it
shows a nice consistence. We can find that the lower end of
strength is obviously affected by the parameter $l$ and $m$ and
departure from power-law, while the upper end is still distributed
as power-law. In the section $\S$\ref{Purelyweight}, we will
compare the analytical results with that of computer simulation in
Fig(\ref{Weight_Leffect}). They are also consistent very well.

\section{\label{Simulation}Numerical results}

\subsection{Degree-driven Model}
\begin{figure}
\includegraphics[width=7cm]{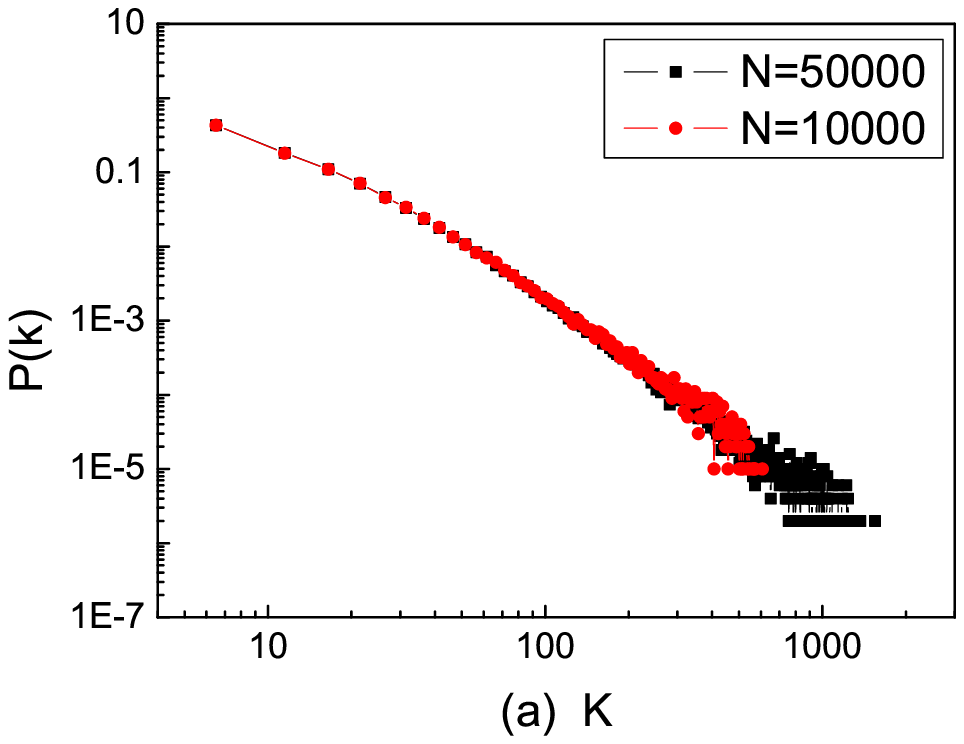}\includegraphics[width=7cm]{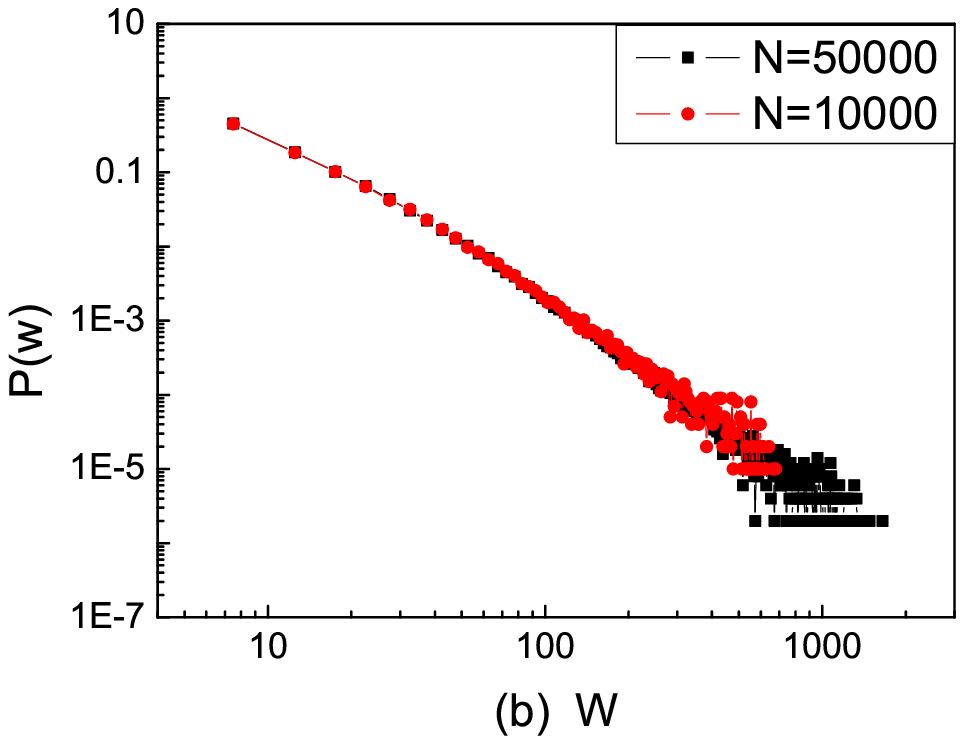}
\center \includegraphics[width=7cm]{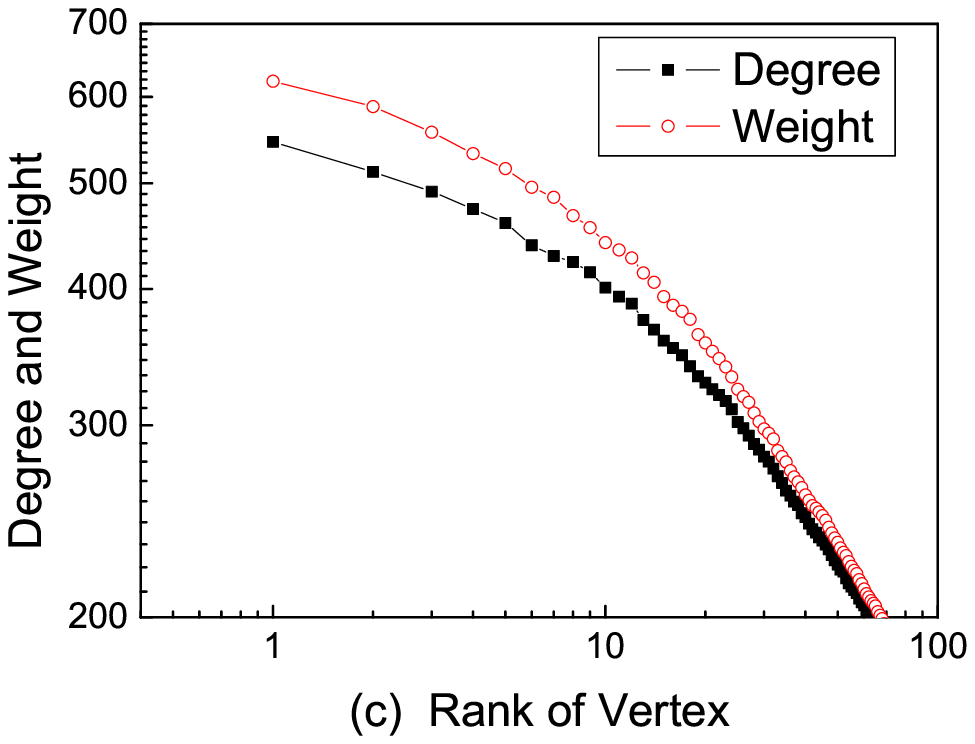} \caption{Degree
(a) and vertex weight (b) distribution from degree-driven model.
They are both power-law distribution with almost the same
exponent, $\gamma=-2.7$. The difference between these two
distributions in upper tail is shown in (c) by Zipf plot. But most
points are in the lower region. In this simulation, $n_{0}=10$,
$m=5$, $l=1$.} \label{BAvweight}
\end{figure}

\begin{figure}
\includegraphics[width=7cm]{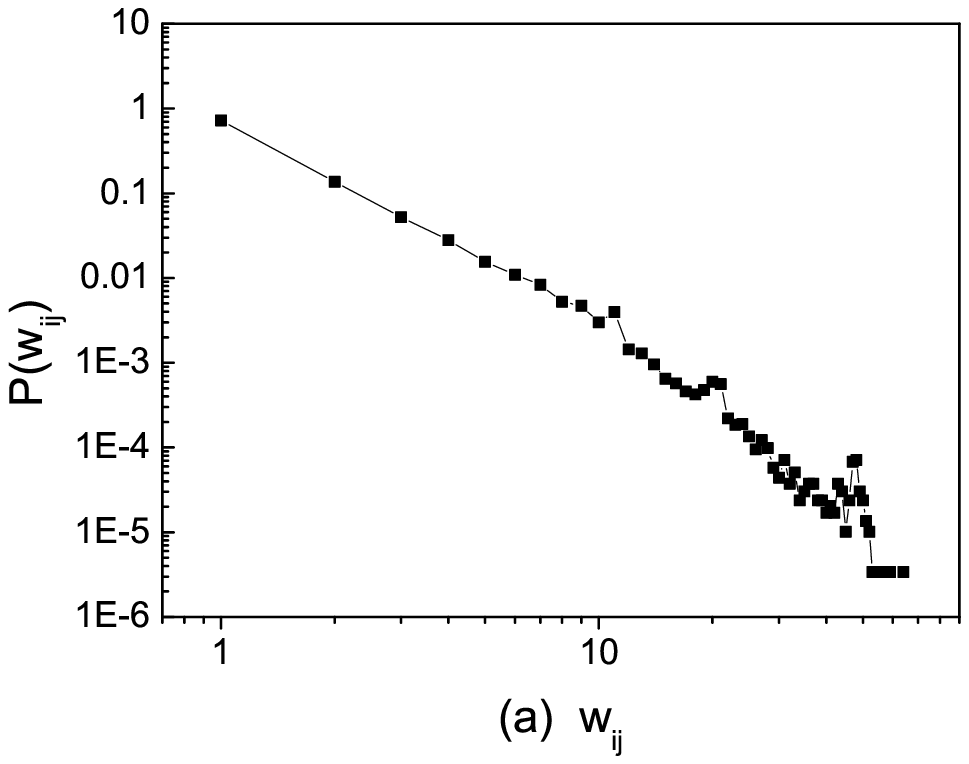}\includegraphics[width=7cm]{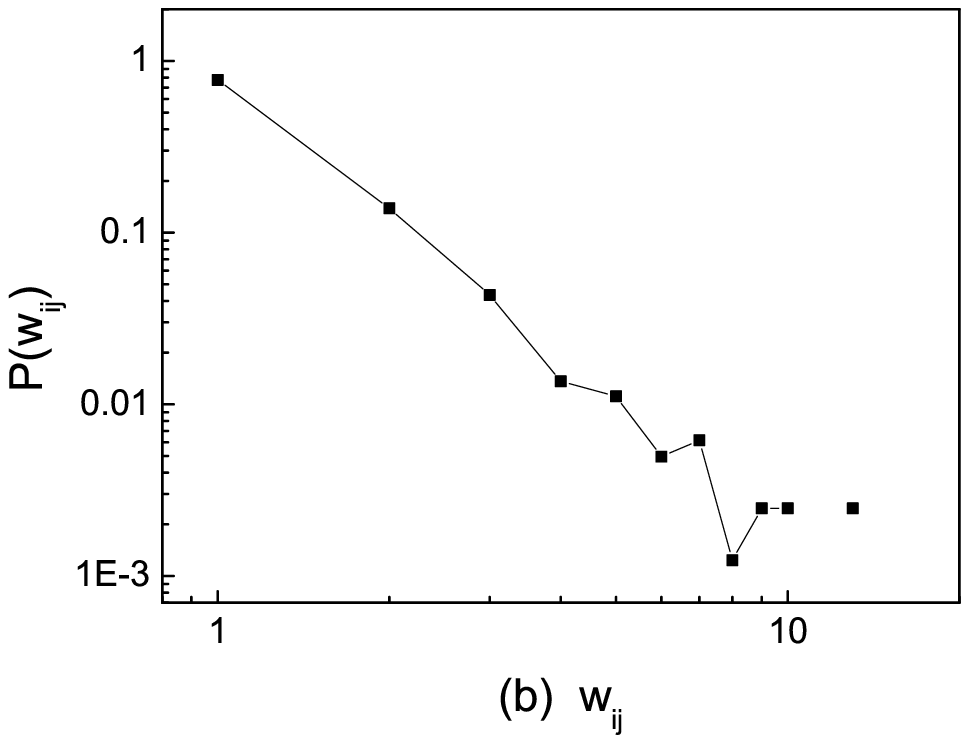}
\center \includegraphics[width=7cm]{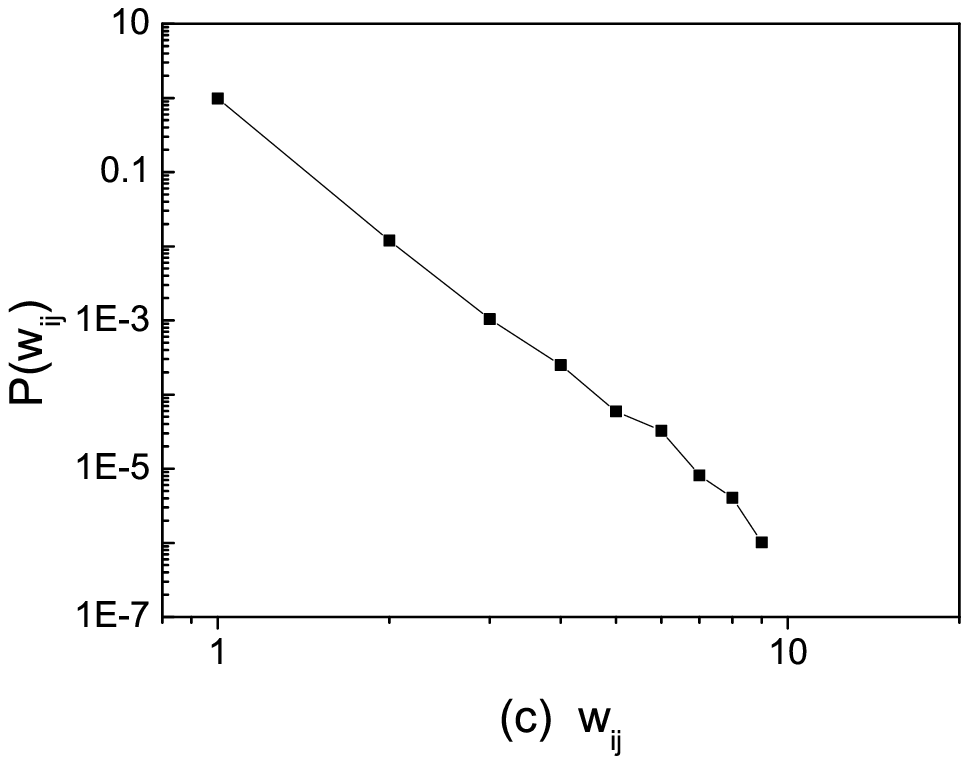} \caption{Edge weight
distribution. Empirical results from Newman's data of scientific
collaboration networks (a) and collaboration network of
Econophysists (b). Simulation result of the model is shown in
(c).} \label{Linkweight}
\end{figure}

First, we consider a special case, $p=0$ and $\delta=0$. In this
case, our model is fairy similar with BA model, except now,
besides the new vertex added in, the old vertices can also be
activated. This assumption has been used already in several
evolving models especially for the modelling of cooperation
networks\cite{Actor,Barabasi,Michele}. Another difference between
this case and BA model is that the reconnection of link is allowed
and recorded. Later on, it will be converted as the weight of
link. Fig(\ref{BAvweight}) shows the typical behavior of degree
and vertex weight distribution. The weight distribution of links
obeys also power law as shown in Fig(\ref{Linkweight}). These
results are consistent with the typical result from empirical
studies qualitatively, such as distribution of vertex weight for
airline networks\cite{W.Li,Barrat1,Bagler} and collaboration
networks\cite{Barrat1}.

\begin{figure}
\includegraphics[width=7cm]{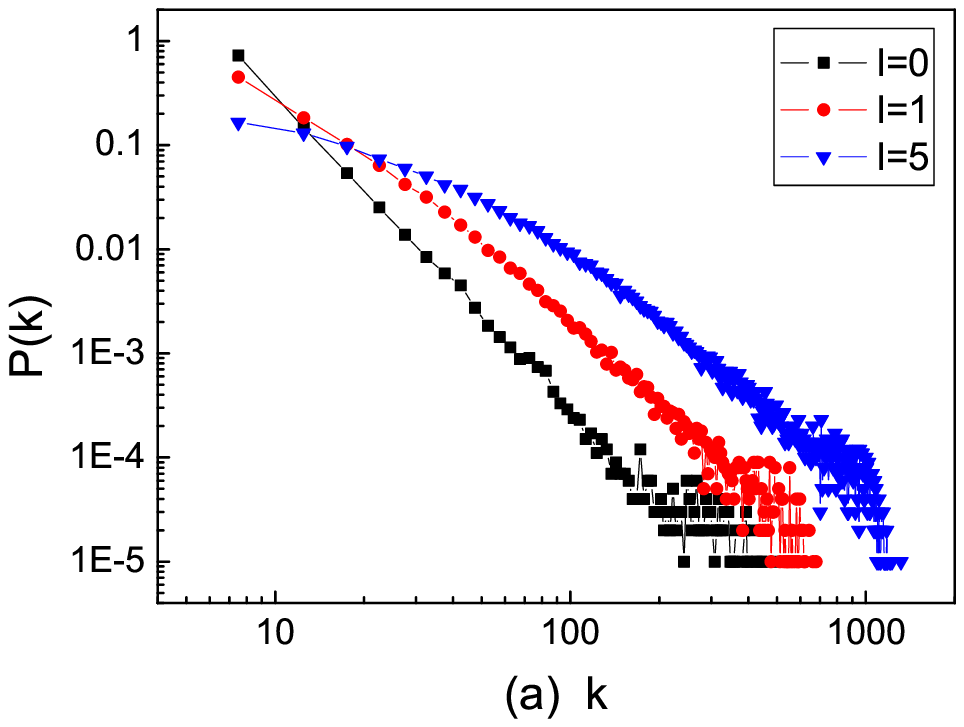}\includegraphics[width=7cm]{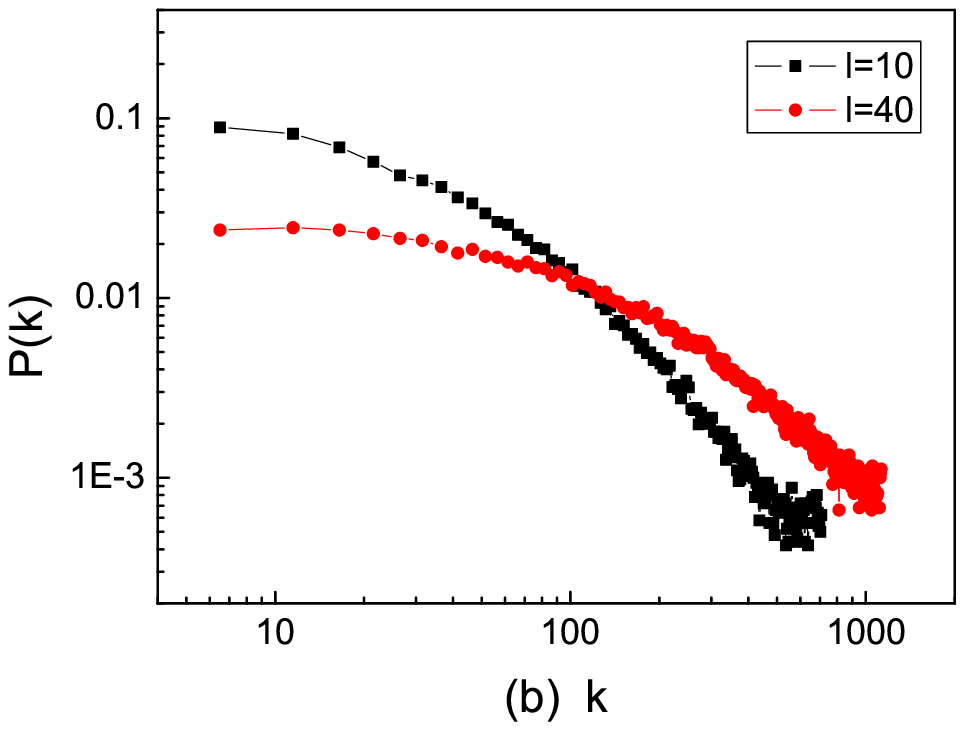}
\caption{(a) Degree distribution affected by the parameter $l$.
(b)Larger $l$ ($l=10, 40$) makes the power-law behavior lose at
the lower end. Other parameters are $n_{0}=10$, $m=5$.}
\label{leffect}
\end{figure}

Compared with BA model, the introduction of parameter $l$ is new,
so how $l$ will effect the behavior of the model? One limit
situation is when $l = 0$. All the contributions to the weight
come only from the new vertices. So our model comes back to BA
model except some new links may be repeated. The degree
distribution is the same as BA model. The vertex weight
distribution is almost the same as degree distribution but there
is no power-law distribution of edge weight at all. The increasing
of $l$ will affect the degree distribution. The lower end will
departure from the power-law distribution but show the "droop
head" shape observed in many empirical studies. Another limit
situation is $l\gg1$. In that case, the increase of internal links
has much more effects on the network evolution compared with the
growth of the network. The network will lose the power-law
behavior in the lower end, although in a quite large domain of
$l$, the power-law behavior of degree and weight distributions are
robust, especially in the upper end. The effect of $l$ is shown in
Fig(\ref{leffect}).

All the results are the average of 10 simulations for different
realization of networks under the same set of parameters. The
network sizes are all reach 10000 nodes. We have compared the
distribution with that of the network with 50000 nodes. They are
almost the same so that a network with 10000 nodes can give us a
nice description for asymptotic distribution.

\subsection{\label{Purelyweight}Purely weight-driven model}
Now we assume that the vertex weight plays the most significant
role in the evolution so that $p = 1$ and $\delta = 0$. This means
the scientists choose their cooperators according to the weight,
instead of focusing on degree. Therefore, weight is the
fundamental character of vertices. Intuitively, the meaning of the
degree looks like the extensiveness of the working style while the
weight considering both extensiveness and intensiveness. So it's
not very surprised that weight can unconsciously be used as a
scale to attract more cooperators. In fact, the same idea of this
weight-driven mechanism has been used in Barrat's
paper\cite{Barrat}. As we mentioned in introduction, the only
difference between Barrat's model and this model is the evolution
mechanism of weight. In Barrat's model, it evolves directly by a
phenomenological rule as a $\delta$-increase, while in our model,
it evolves indirectly through the connecting times $T_{ij}$.

\begin{figure}
\includegraphics[width=7cm]{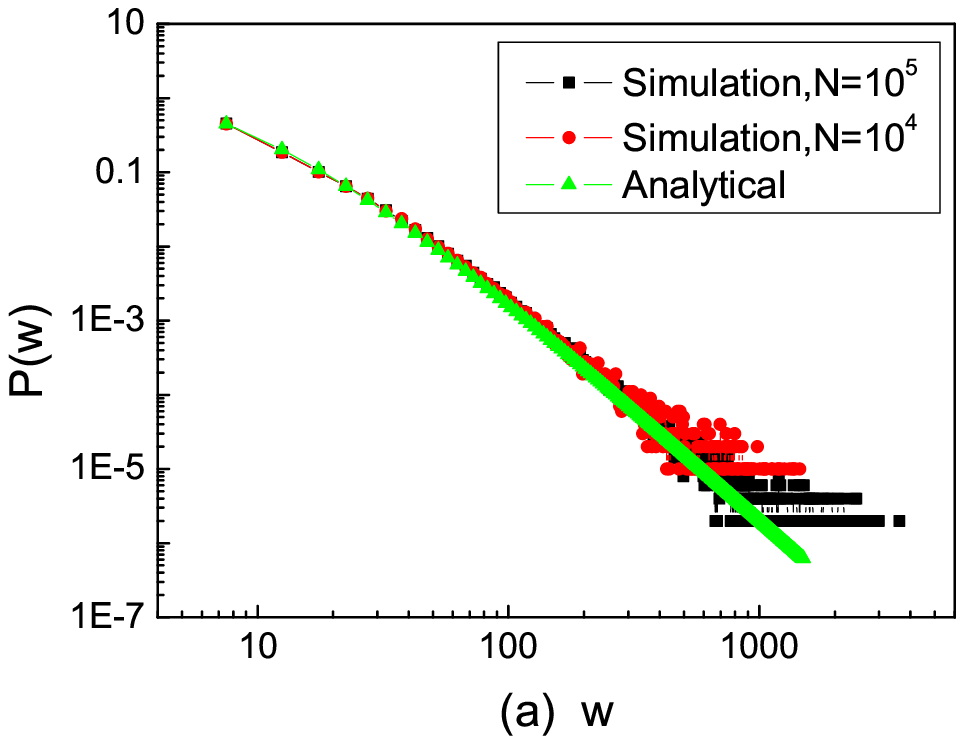}\includegraphics[width=7cm]{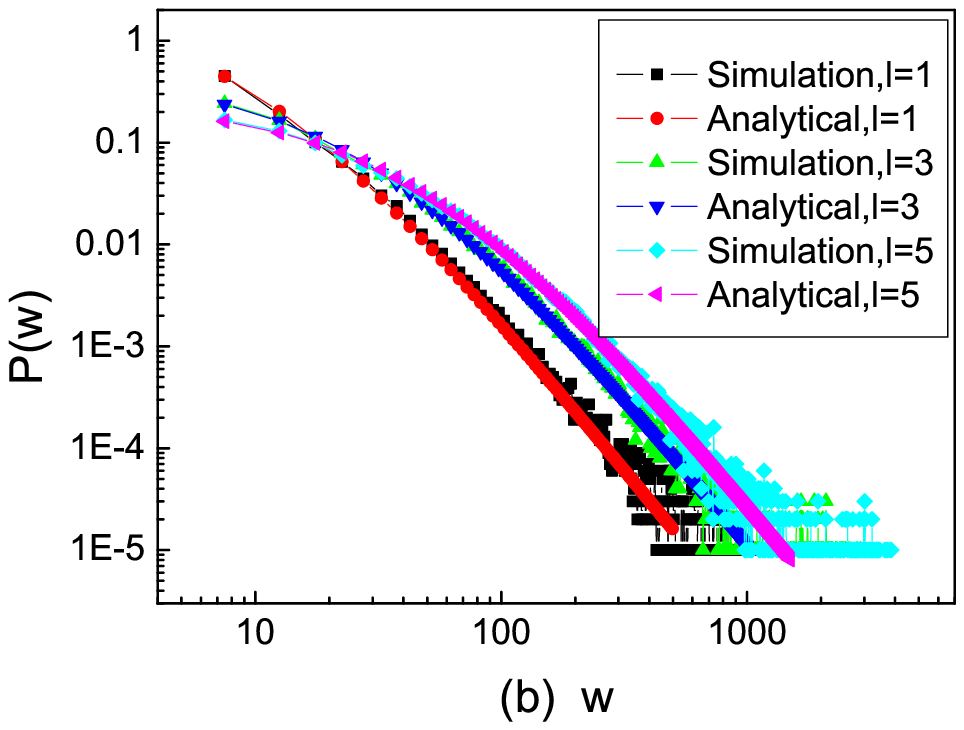}
\caption{Vertex weight distribution for the purely weight-driven
model: (a)comparison of the simulation results with analytical
results; (b) the effects of different values of $l$. Other
parameters are $n_{0}=10$, $m=5$.} \label{WeightLeffect}
\end{figure}
\begin{figure}
\includegraphics[width=7cm]{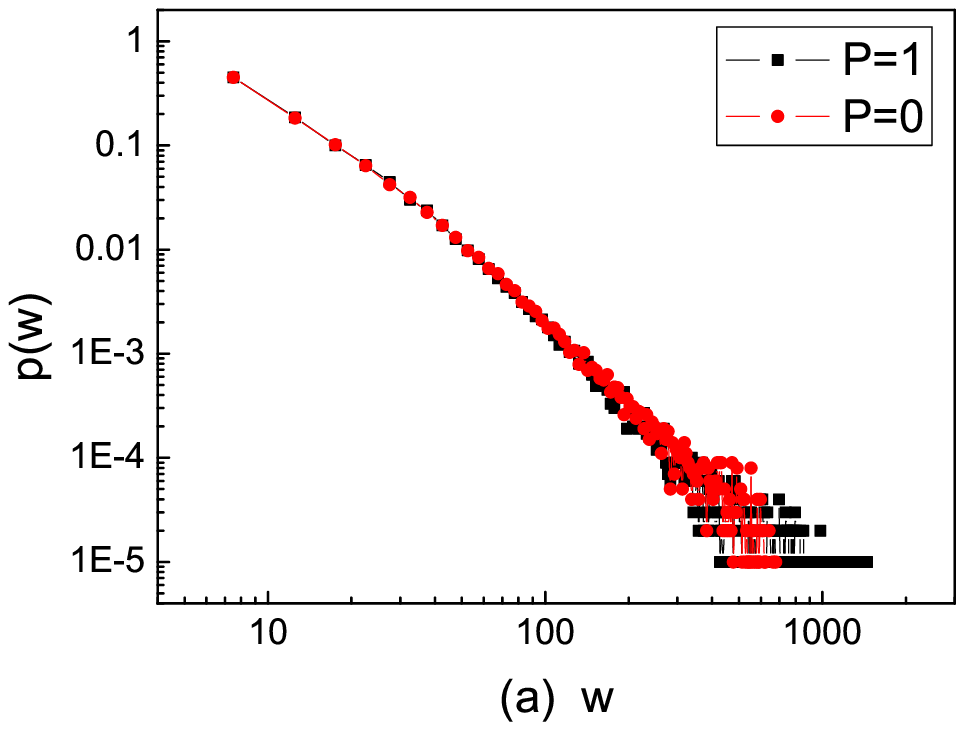}\includegraphics[width=7cm]{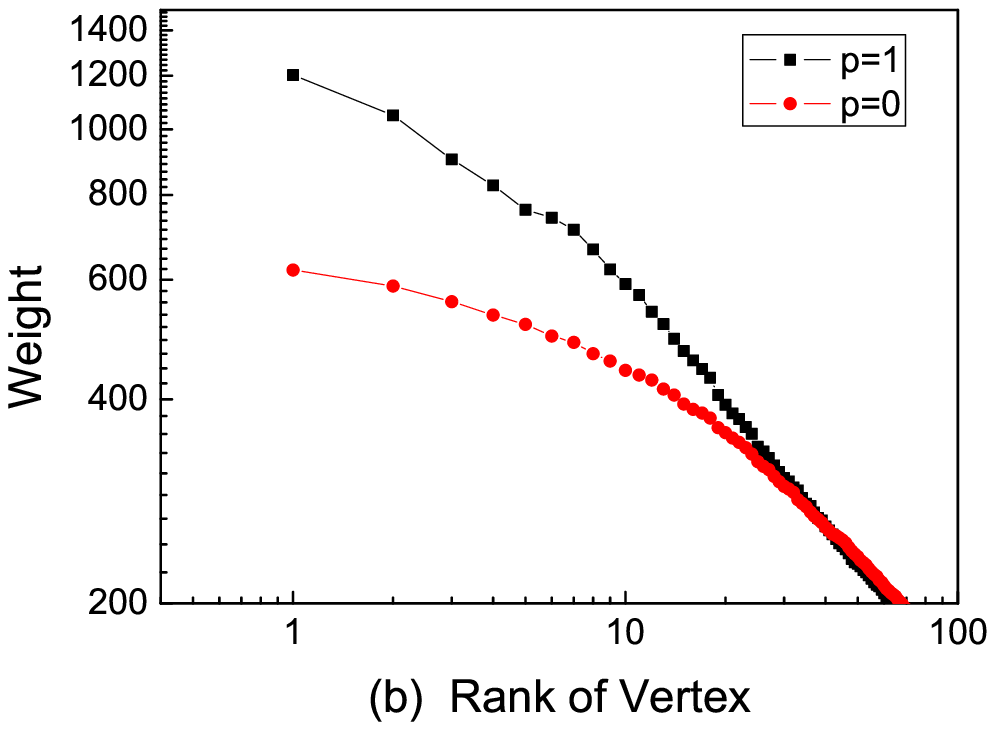}
\caption{Comparison between purely degree-driven and purely
weight-driven models on weight distribution. We could see that
those two models behavior very similarly (a) except for the upper
tail in Zipf plot (b). The parameters for two simulations are the
same: $n_{0}=10$, $m=5$, $l=1$.} \label{compare}
\end{figure}

The numerical results are given in Fig(\ref{WeightLeffect}). They
are consistent with the theoretical analysis. It has been found
that those two limit cases seems have the similar qualitative
behavior. However, as we mentioned in the introduction, the
difference between those two cases, and the conclusion that which
one behaviors better, implies the answer to the question that
which quantity is the more fundamental one between degree and
weight. Or put it in another way, should weight be a high-level
quantity defined by degree, betweenness, whatever the basic
network quantities, or directly from event represented by the
network? Therefore, an conclusion about this comparison is
essential for this issue. However, so far those two models under
the limit cases provide the similar behavior. In the next section
$\S$\ref{extend}, when the models are extended onto multilevel
relationships to do a comparison between models and empirical
results, at first we extended both those two limit cases. Both of
them provide consistent behavior with the empirical results. After
the similar results are found, only weight-driven model is
compared with empirical results further. Of course, one can even
try to study the behavior for a general $p$ besides those two
limit cases. However, since we have no way to make significant
difference between the special cases, no further investigation
about the general model has been done in this paper. To compare
those two models with more empirical results and find the
significant difference between them is really valuable for network
analysis.

\subsection{Clustering structure when $\delta\neq0$}

The mechanism represented by the $\delta$ term in our model has
taken the local information of the network into account. When a
connection is built up by the active vertex, the probability of a
vertex being chosen as the end vertex is higher if it has the
closer relation with the attempting one. Therefore, hopefully,
this will increase the clustering coefficient. In the original BA
model, this mechanism is neglected. So it's interesting to just
keep the degree term and the $\delta$ in our model, and explore
whether such mechanics increase clustering coefficient or not.

\subsubsection{$\delta$-mechanism applied onto model of non-weight networks}

For the non-weighted networks, every edge has the weight 1. The
$l_{ni}$ in $\delta$ term is the similarity distance, which is the
reciprocal of the shortest distance between vertex $n$ and $i$.
For the first link from new attempting vertices, just the BA rule
of preferential attachment is applied, but for the links
afterwards and the links from old vertex, the end point is
determined preferential by both its degree and closeness with
starting point. In Fig(\ref{cluster}), the simulations show this
$\delta$ mechanism significantly increases the clustering
coefficient while the power-law distribution of degree still
holds.

\subsubsection{$\delta$-mechanism in weighted network}

\begin{figure}
\includegraphics[width=7cm]{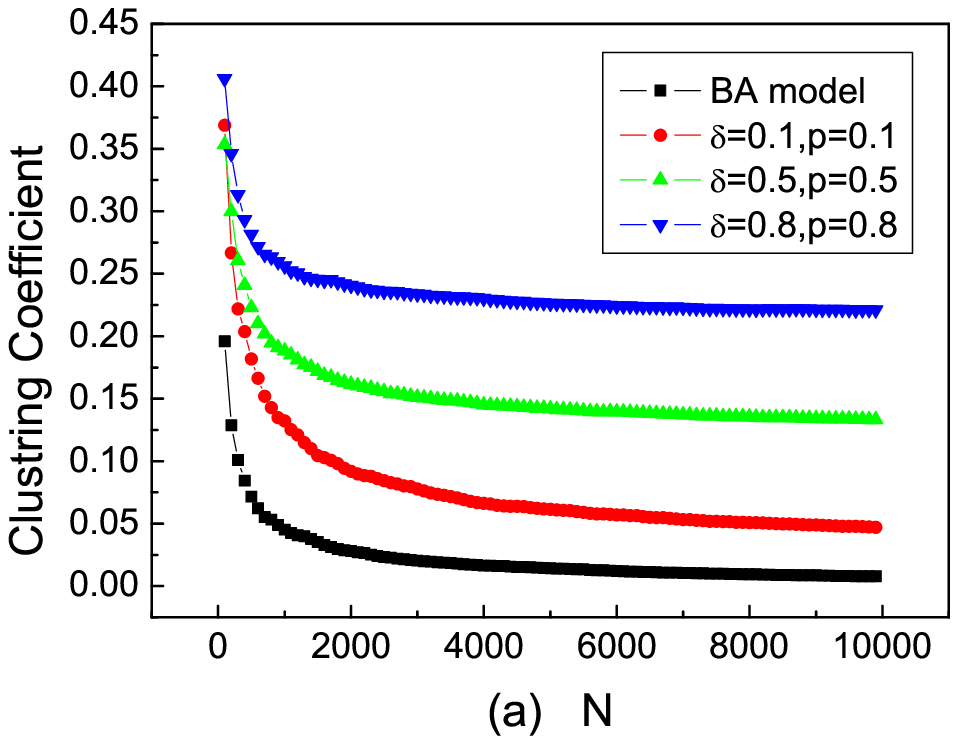}\includegraphics[width=7cm]{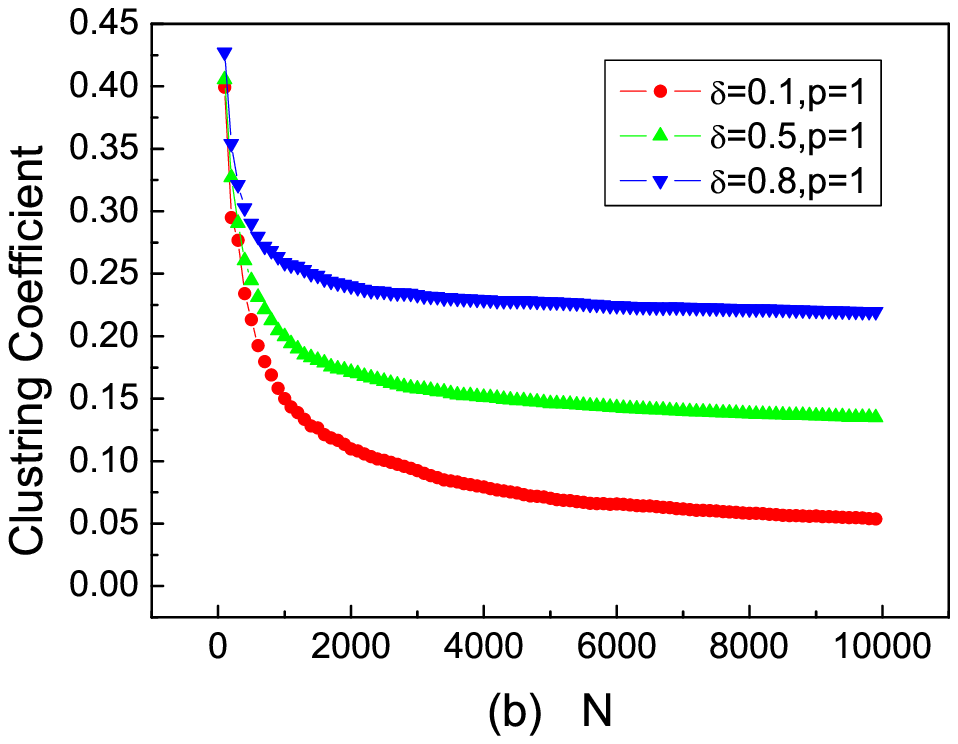}
\caption{Clustering coefficient for degree-driven (a) and
weight-driven (b) model. Parameters are $n_{0}=10$, $m=5$, $l=1$.}
\label{cluster}
\end{figure}

For the weighted networks, $l_{ni}$ in $\delta$ term is the
similarity distance as mentioned before. In this case, just for
simplicity, we consider the purely weight-driven model, which
means $p = 1$, $\delta \neq 0$ in equ(\ref{prob}). Its effects on
clustering coefficient are also shown in Fig(\ref{cluster}).
The clustering coefficient under this mechanism reaches a stable
value after a period of evolution. We can see that under some
value of $\delta$, for example, $\delta=0.8$, the clustering
coefficient is near $0.25$, which is much higher than BA model,
and even comparable with empirical
results\cite{Newman1,Li,Barabasi} and results from other
models\cite{Holme,Davidsen}.

Since we have already shown that the $\delta$-mechanism can
increase the clustering coefficient, from now on, we will focus
only on the comparison between simulation and empirical results on
the distribution of degree and weight, instead of on the
clustering coefficient. So in the following discussion, $\delta$
is set to be $0$ again.

\section{\label{extend} Extended model and comparison with empirical results}
In the real world, relations of nodes usually are more than one
levels and different relations have different contributions to the
weight of link. For instance, in the empirical analysis in
\cite{Fan, Li}, we consider both co-authorship and citation as the
ways of scientific idea transportation with different
contributions. Even the worse is that citation is a directed
network. So in order to compare the results from our models with
the empirical studies, we must extend our current model into a
multilevel directed networks model. There are two kinds of
connecting times $T^{\mu}_{ij}$, where $\mu = 1,2$ refers to
co-authorship and citation respectively. Here the relation between
connecting times and the link weight is given by a $\tanh$
function. The reason we prefer the $\tanh$ function in empirical
studies is that, first, it has the saturation effect, which makes
the contribution less for larger connecting times; second, it
normalizes the maximum value to $1$, which is the usual strength
of edge in non-weight networks. So the two $T^{\mu}_{ij}$ are
converted into a single weight by
\begin{equation}
w_{ij}=\frac{1}{2}\sum_{\mu}\tanh\left(\alpha_{\mu}T^{\mu}_{ij}\right),
\label{weight2}
\end{equation}
so that $w_{ij}$ is normalized to $1$. And the probability
distribution to chose the end vertex is consistently transformed
as
\begin{equation}
\Pi_{n\rightarrow i} =
\sum_{\mu}p^{\mu}\frac{w_{i}}{\sum_{j}w_{j}},
\end{equation}
while $\sum_{\mu}p^{\mu} = 1$. Or in degree-driven model,
\begin{equation}
\Pi_{n\rightarrow i} =
\sum_{\mu}p^{\mu}\frac{k_{i}}{\sum_{j}k_{j}}.
\end{equation}
After vertex $i^{*}$ are chosen as the end vertex of a relation
$\mu$ between $n$ and $i^{*}$ according to above probability
distribution, the connecting time evolutes as
\begin{equation}
T^{\mu}_{ni^{*}}\left(t+1\right) = T^{\mu}_{ni^{*}}\left(t\right)
+ 1.
\end{equation}
For $\mu=1$, after that we need to set $
T^{1}_{i^{*}n}\left(t+1\right) = T^{1}_{ni^{*}}\left(t+1\right)$.
For $\mu=2$, we skip this step. From its definition
equ(\ref{weight2}), the weight here is an integrated variable.
This implies those two events can be triggered by each other, not
developed separately.

\begin{figure}
\includegraphics[width=7cm]{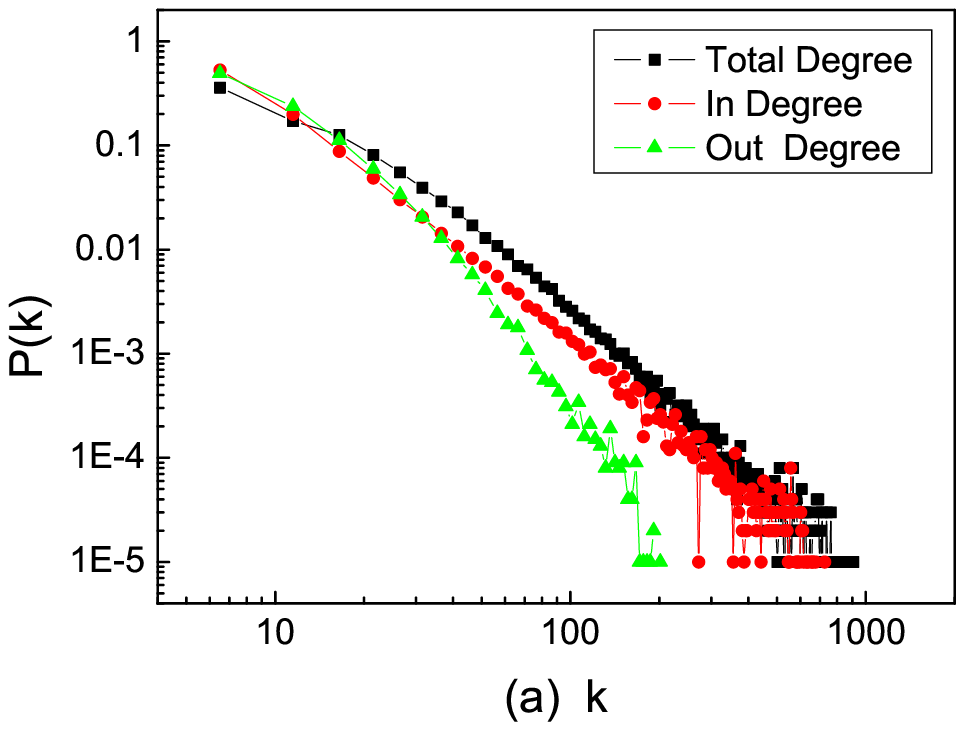}\includegraphics[width=7cm]{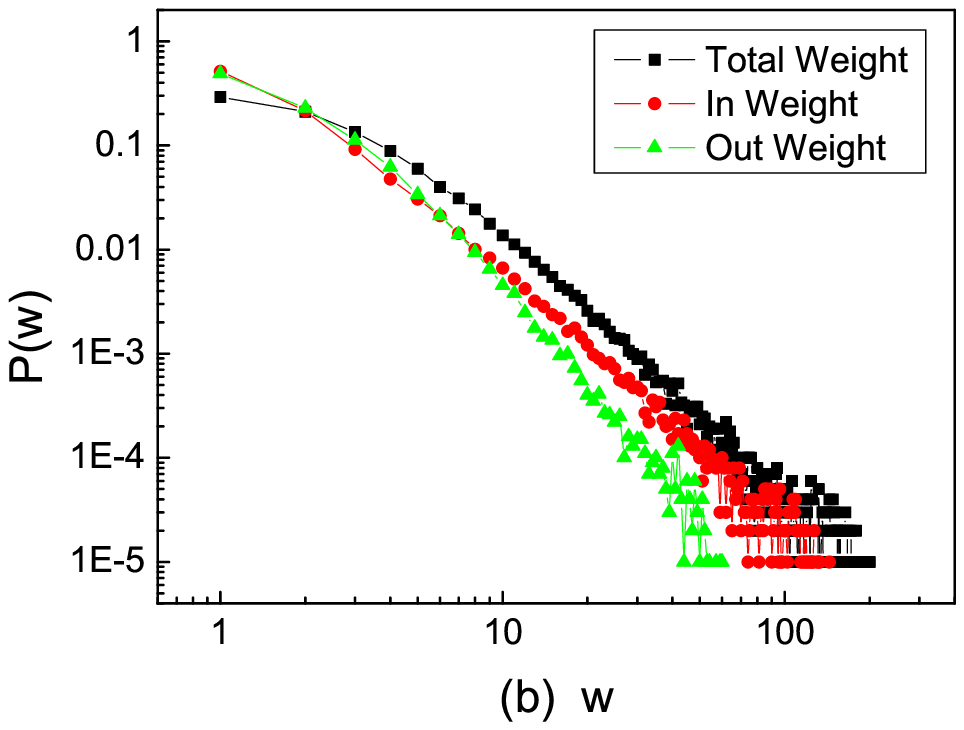}
\includegraphics[width=7cm]{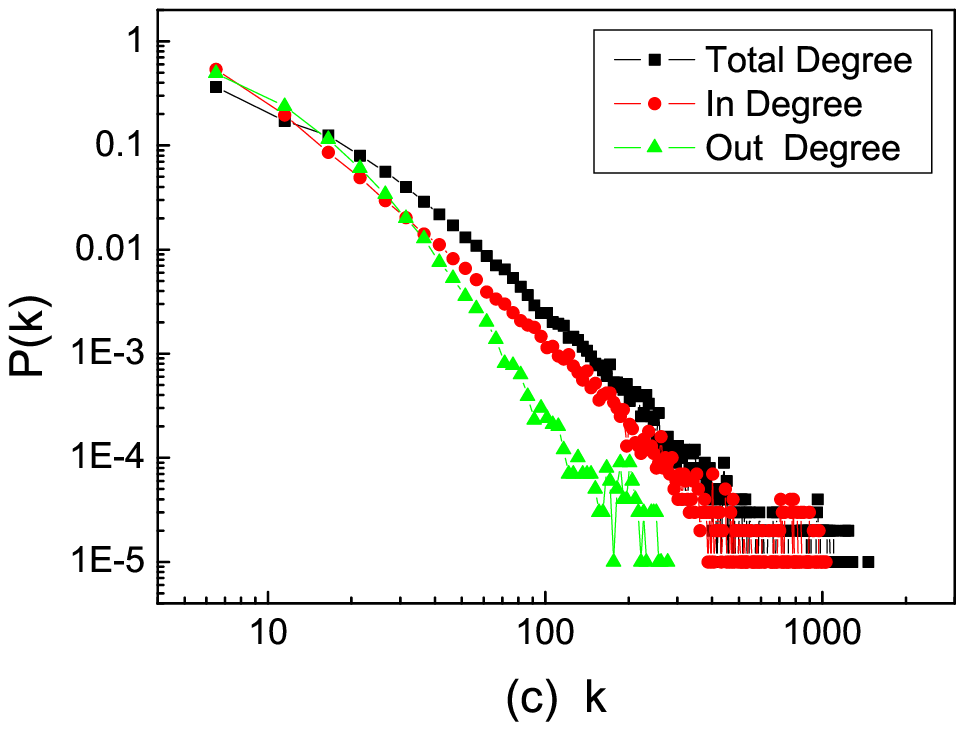}\includegraphics[width=7cm]{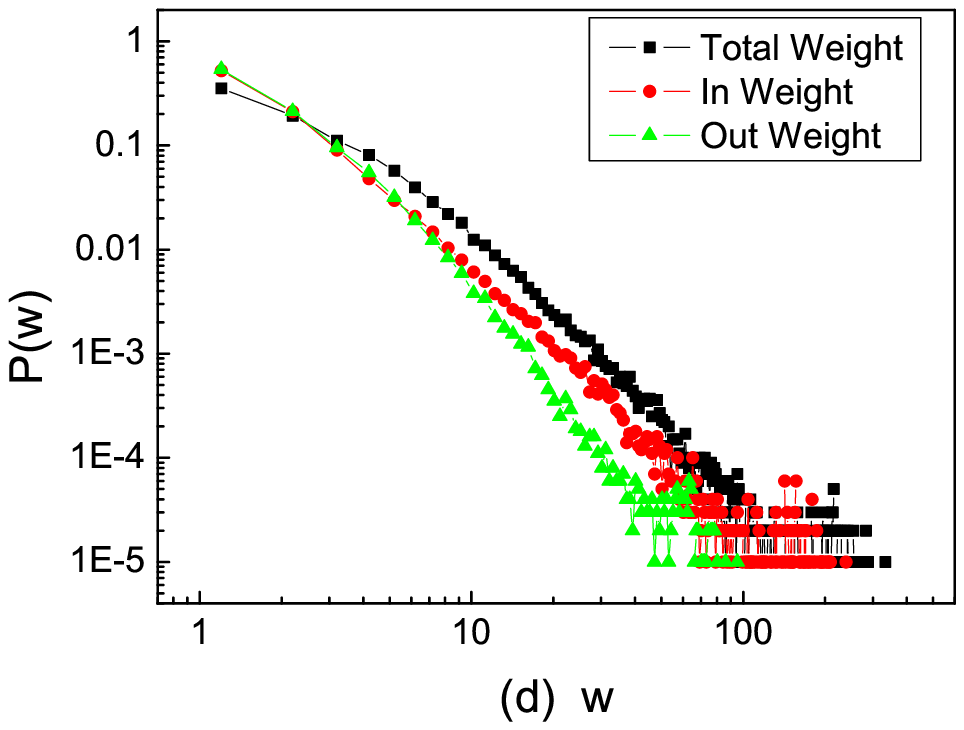}
\caption{Typical distributions of degree and weight (the three
curves are In, Out and Total respectively) from pure degree-driven
model (a), (b) and pure weight-driven model (c), (d). Again, the
pure weight-driven model gives very similar results.} \label{IO}
\end{figure}

As it will be shown in Fig(\ref{IO}), we have not found any
significant difference between those two models, so later on, when
we compare models with empirical results, only weight-driven model
are used there. As explained in the introduction section
$\S$\ref{Introduction}, measuring the role of weight by
evolutionary models is one of the goals of our research which has
not been achieved so far. We hope more comparisons between the
behaviors of those two models and more empirical results will give
an conclusive answer for this question.

For directed network, the degree is divided into three quantities:
out degree, in degree and total degree. For example, the in degree
$k^{in}_i$ is the sum of edges ending at vertex $i$, that is
$k^{in}_i=\sum_j{sign\left(w_{ji}\right)}$.  The out degree and
weight are calculated similarly and the total degree and weight
are the sum of in and out. The same situation happens to vertex
weight, so there are out weight, in weight and total weight of
vertices. From the simulation results, we can find that the total,
in, and out degree and weight are all of power law distribution,
as shown in Fig(\ref{IO}).

The more important comparison we want to do is between simulation
and empirical results, especially on link and vertex betweenness,
because they are global properties related with the whole
structure of the networks. The link and vertex betweenness and
their distribution could be gotten from the set of effective
pathes between any two nodes. The average distance $d$ is defined
as before as,
\begin{equation}
d = \frac{1}{N(N-1)}\sum_{ij} d_{ij}
\end{equation}
in which, $d_{ij}$ is the similarity distance of an effective path
between vertex $i,j$, the larger the closer and equals to $0$ if
no path exists. In fact, the above formula is not exactly the same
as the one for non-weighted networks. First, because of the
direction of edges, the number of total edges are now
$N\left(N-1\right)$ instead of $\frac{N\left(N-1\right)}{2}$.
Second, the algorithm to search for such $d_{ij}$ is slightly
different with the usual shortest path in non-weighted networks.
One way to make use of the shortest path algorithm is to transform
the similarity into dissimilarity weight, so that the shorter the
closer, and then use the usual shortest path algorithm to find all
the distance. After that, transform it back into similarity
distance. However, this is just an algorithm problem, has nothing
to do with the structure analysis.

In order to check the model, we compare the results with empirical
results from Econophysicists network, which has mostly been given
in \cite{Fan,Li}. We compare the distribution of quantities of
Econophysicists networks with numerical simulations, such as
degree, vertex weight, vertex betweenness and link betweenness. It
is interesting that the results are consistent well. It seems that
the model reveals some basic mechanisms of the evolution of
collaboration network. The parameters we used here for this
comparison are $l = 1, p^{1} = 0.2, p^{2} = 0.8$.
\begin{figure}
\includegraphics[width=7cm]{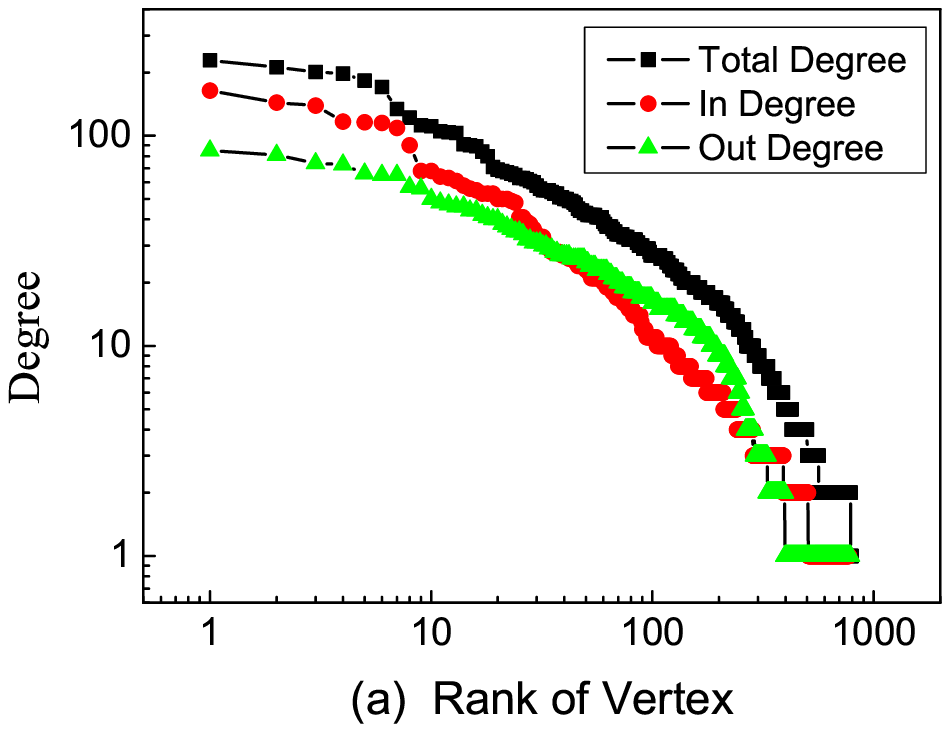}\includegraphics[width=7cm]{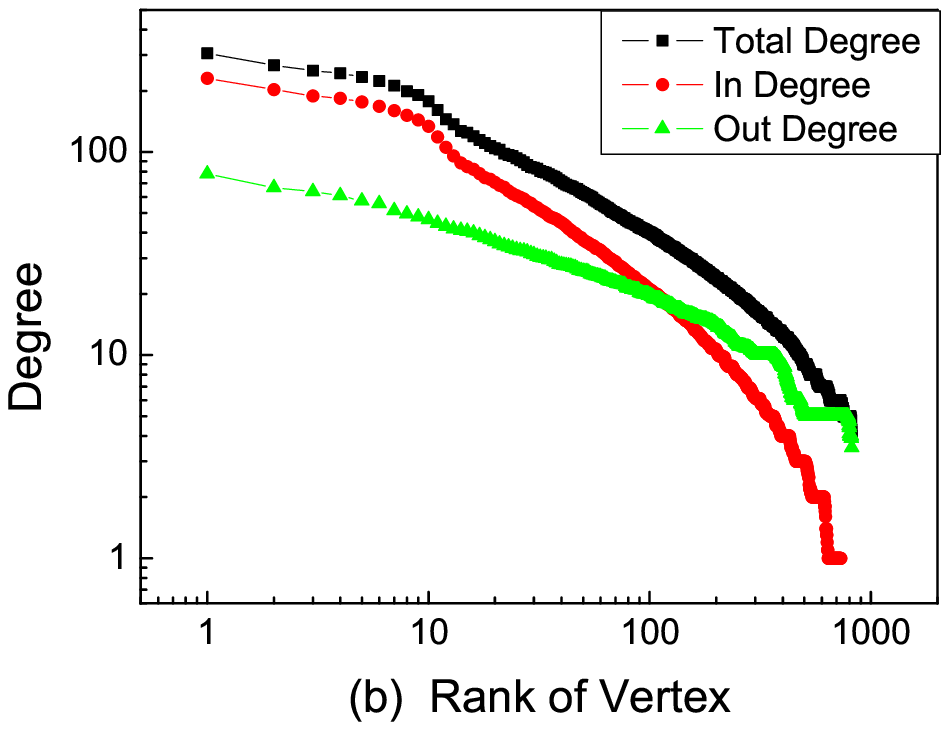}
\includegraphics[width=7cm]{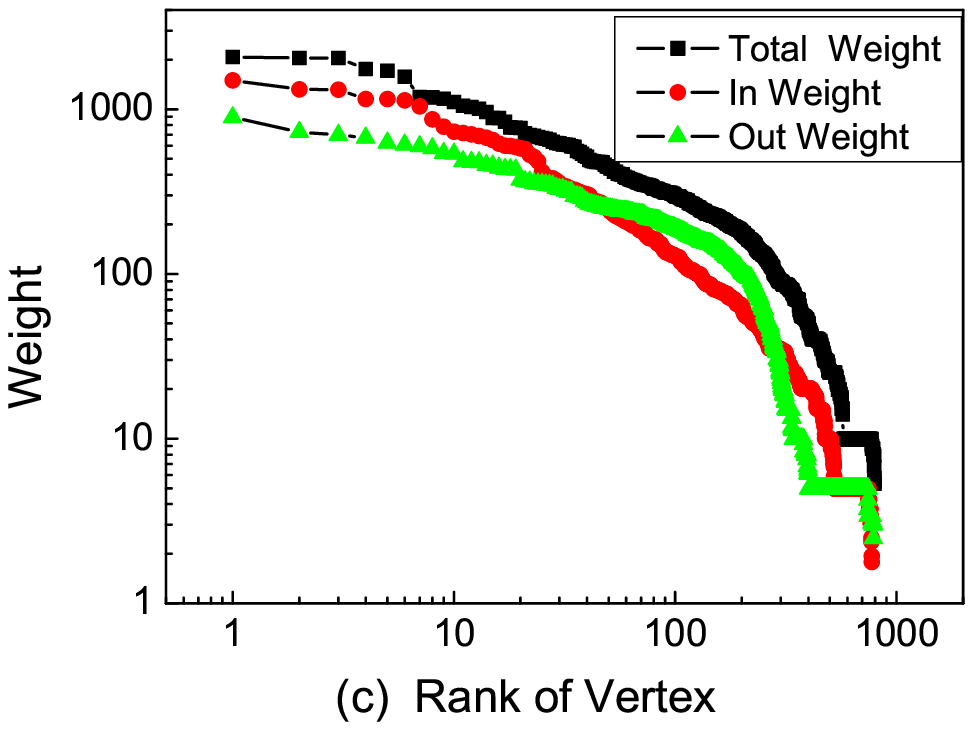}\includegraphics[width=7cm]{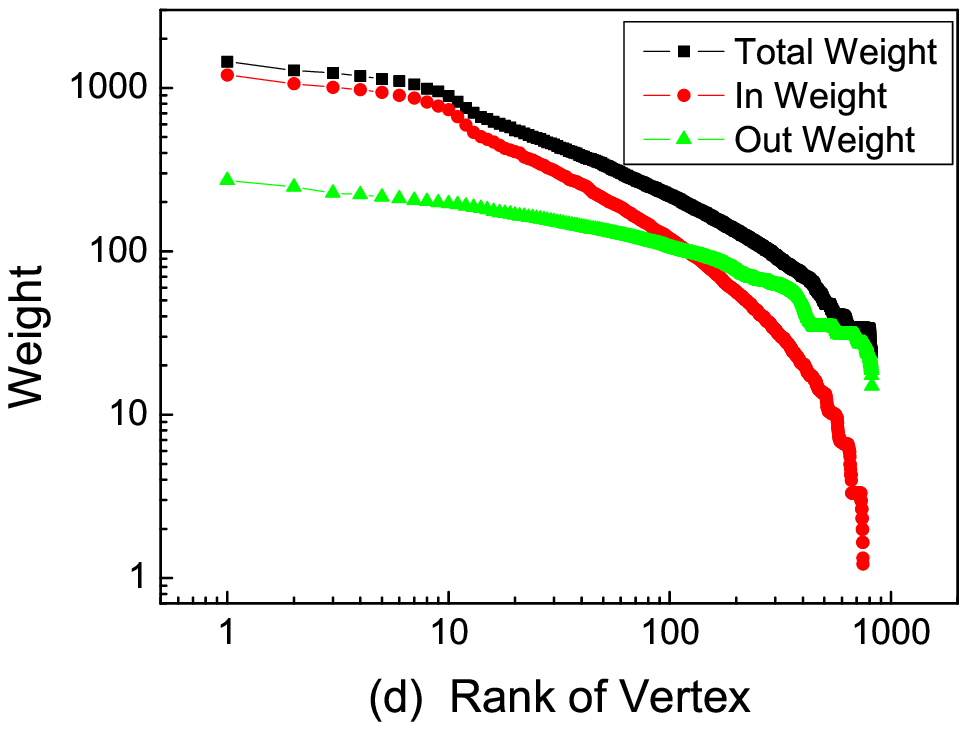}
\caption{Zipf plot of degree and weight from empirical studies and
simulations. (a) and (c) are the empirical results of degree and
weight distribution. (b) and (d) are the simulation results of the
model. The model is simulated under the parameters: $n_{0}=10,
m=5, l=1, p^{1}=0.2, p^{2}=0.8$. $\alpha^1=0.7, \alpha^2=0.3$ in
equ\ref{weight2}. The size of simulated network is $N=819$ that is
the same as empirical studies.} \label{DRS}
\end{figure}
\begin{figure}
\includegraphics[width=7cm]{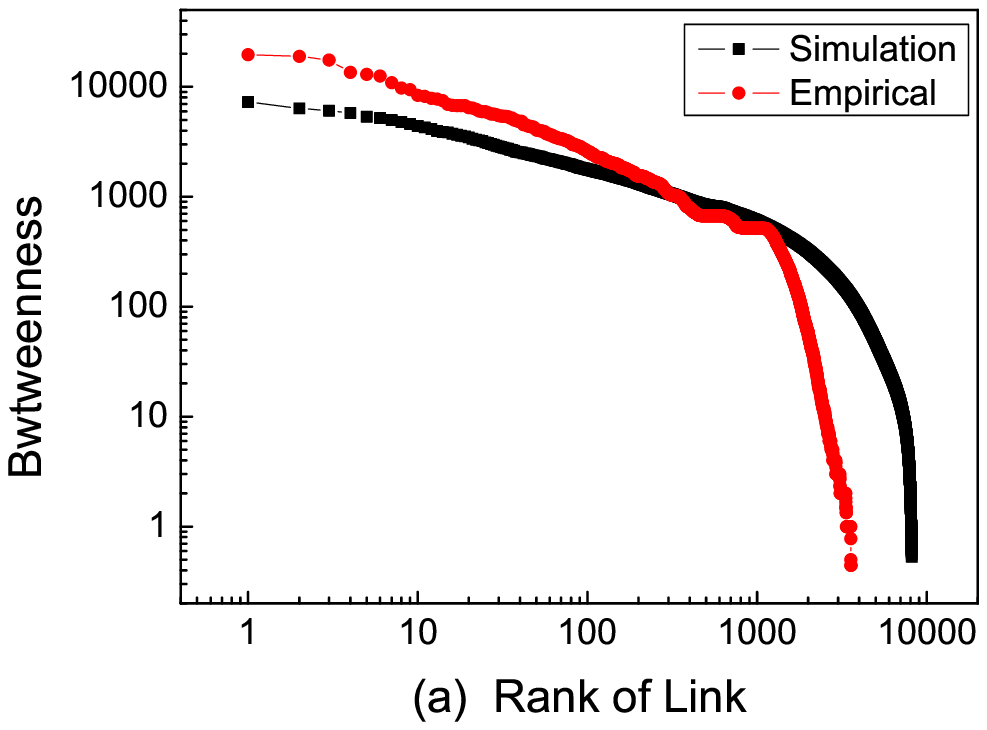}\includegraphics[width=7cm]{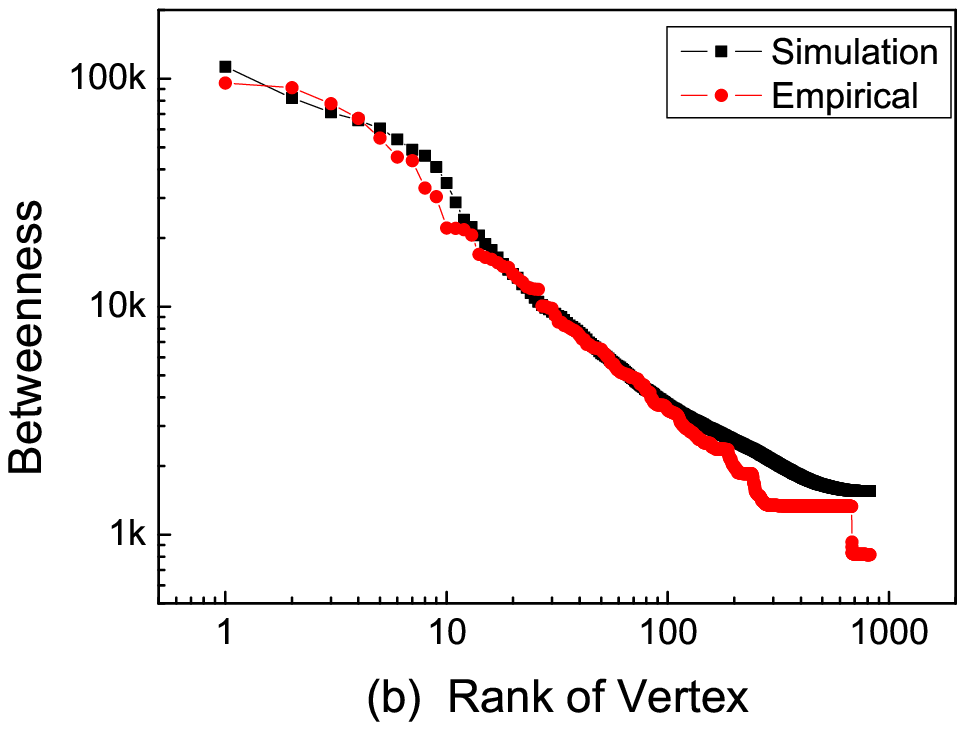}
\caption{Zipf plot of betweenness for empirical studies and
simulations. (a) Link betweenness.  (b) Vertex betweenness.}
\label{BRS}
\end{figure}

\section{Concluding Remarks}
In this paper, we presented an evolutionary model for weighted
network, which integrates the contributions from both new vertices
and old vertices. The two mechanisms, degree-driven and
weight-driven preferential attachment are discussed, and both show
a good consistence with empirical results from network of
econophysicists. Also a new mechanism, named as local-path-related
preferential attachment, which makes use of some locality
information is introduced here to increase the clustering
coefficient of the network. Weight has been assigned to each link
according the connecting times of the link, so the weight of link
changes as network evolutes. Including the behaviors from the old
vertices, recording all the connecting times and converting them
into weight, are the most essential steps in our model. The way to
incorporate locality information into network evolution by the
$\delta$-mechanism is also one point of this paper.

However, one of the most important tasks of this paper, which is
to determine the role of weight and comparing it with the role of
degree, has not been done yet. Although the comparison so far
could not distinguish the degree-driven model and weight-driven
model, we hope further comparison with empirical results will give
a conclusive answer for this question.

\section{Acknowledgments}
We thank Prof. Dinghua Shi, Dr. Yiming Ding, Dr. Qiang Yuan for
their helpful discussions and comments on this paper. The work is
partially supported by NSFC under the grant No.70431002 and
No.70471080.

\bibliography{apssamp}

\end{document}